%% file: main.tex
\documentclass{dais} 
\usepackage{xcolor}

\usepackage{enumitem}
\usepackage{amsmath}
\usepackage{listings}
\usepackage{inconsolata}
\definecolor{jevcodegreen}{rgb}{0.228,0.695,0.195}
\definecolor{jevcodepink}{rgb}{0.7539,0.1992,0.4531}
\definecolor{jevcodeblue}{rgb}{0.1914,0.457,0.707}
\definecolor{jevcodegray}{rgb}{0.59,0.59,0.59}
\definecolor{jevcodeyellow}{rgb}{0.9254,0.5529,0.366666667}
\lstdefinelanguage{JevSQL}{
    morekeywords={CREATE,MODEL,ON,PROMPT,PATH,API,SELECT,FROM,WHERE,HAVING,
                  ORDER,GROUP,BY,DESC,ASC,AS,LLM,AGG,OPTIONS,INNER,NATURAL,
                  TABLE,JOIN,AND,OR,WITH,LIMIT,TEMP,VIEW,AVG,COUNT,ROUND,
                  INTEGER,VARCHAR,BOOLEAN,DOUBLE},
    sensitive=true,
    morecomment=[l]{--},
    morecomment=[s]{/*}{*/},
    moredelim=[s][\color{jevcodepink}]{\{\{}{\}\}},
    morestring=**[d]{'},
    morestring=[d]{\\'},
    morestring=**[d]{"}
}
\lstdefinestyle{jevsql}{
    language=JevSQL,
    basicstyle=\footnotesize\ttfamily,
    keywordstyle=\color{jevcodegreen}\bfseries,
    stringstyle=\color{jevcodeblue},
    commentstyle=\color{jevcodeyellow}\bfseries,
    backgroundcolor=\color{white},
    columns=fullflexible,
    keepspaces=true,
    showspaces=false,
    showstringspaces=false,
    showtabs=false,
    breakatwhitespace=false,
    breaklines=true,
    numbers=left,
    numberstyle=\tiny\color{jevcodegray},
    numbersep=5pt,
    frame=none,
    float=htbp,
    captionpos=b,
    aboveskip=3mm,
    belowskip=3mm,
    tabsize=2
}
\title{Prune First, Decide Fast: Scalable Semantic Query Processing with JEVDB}

\author[1]{Zhengle Wang}
\author[1]{Hanxu Yan}
\author[2]{Fuheng Zhao}
\author[1]{Chunwei Liu*}

\affiliation[1]{Purdue Data \& AI System Lab (PDAIS)}
\affiliation[2]{University of Utah}
\correspondence{\email{chunwei@purdue.edu}}

\date{September 2026}

\abstract{
Semantic database systems extend standard SQL with foundation model inference, allowing users to express complex declarative queries over unstructured text and domain-specific attributes. Despite their expressive power, existing semantic engines face an acute scalability bottleneck: invoking autoregressive Large Language Models (LLMs) incurs latencies and monetary costs that are orders of magnitude higher than conventional relational operators. Mitigation strategies such as prompt batching, model cascading, and embedding-based proxy models fail to address the core problem---an architectural \emph{generative impedance mismatch}, wherein multi-billion-parameter models generate open-ended text tokens merely to resolve discrete, typed relational predicates (e.g., Booleans, Enums, and Scores).

In this paper, we present \textbf{JEVDB}, a scalable semantic database system that integrates non-generative, calibrated decision models into relational query processing. JEVDB maps semantic operators directly to single-pass typed decision primitives (\lstinline{Noul}, \lstinline{Choice}, \lstinline{Score}), eliminating autoregressive decoding loops and schema parsing overheads. To minimize expensive model calls, JEVDB combines exact Yannakakis-style semijoin reductions over relational structure with \emph{Semantic Bloom Filters} (SBFs) that conservatively screen candidates across latent semantic edges. Surviving tuples are evaluated through an adaptive three-tier execution hierarchy, selectively escalating only uncertain cases to frontier LLMs.
We evaluate JEVDB on both the public SemBench benchmark and Shelob, a large-scale semantic join workload derived from TPC-DS. On SemBench, JEVDB achieves average latency speedups of $72\times$--$89\times$ (with peak query speedups exceeding $100\times$) and reduces monetary costs by up to $18.7\times$, while delivering equal or superior quality on 18 of 21 queries. On Shelob workloads scaling up to 540K candidate pairs where existing baselines time out, JEVDB completes all queries with $95.7\%$--$97.5\%$ mean F1, pruning $87.4\%$ of candidates before model invocation. An interactive query simulator, source code, and benchmarks are previewed at \url{https://jevdb.org}.

}

\begin{document}

\maketitle

\input{01intro}

\input{02related}

\input{03method}

\input{04eval_tmp}

\input{05conclusion_zhengle}

\bibliographystyle{ACM-Reference-Format}
\bibliography{ref}
\end{document}

%% file: 01intro.tex
\section{Introduction}
\label{sec:intro}

Recent breakthroughs in foundation models and Large Language Models (LLMs) have opened transformative opportunities for data analytics, enabling automated reasoning over unstructured text, multi-modal documents, and complex domain concepts~\cite{saeed2023querying, zhao2024hybrid, info_ret_2024,llm_survey_2025, liu2025variable}. To bring these capabilities directly to relational data practitioners, recent data management research has introduced declarative semantic database systems~\cite{palimpzestCIDR, lotus_2024, flock_2025,liskowski2026cortex}. Systems such as Palimpzest~\cite{palimpzestCIDR, abacus}, LOTUS~\cite{lotus_2024}, and DocETL~\cite{docetl_2025} augment standard Structured Query Language (SQL) with first-class semantic operators, including semantic selections ($\sigma_{\text{sem}}$), joins ($\bowtie_{\text{sem}}$), projections ($\Pi_{\text{sem}}$), and aggregations ($\Gamma_{\text{sem}}$). By unifying relational algebra with model inference, semantic databases allow users to express complex analytical queries declaratively, inheriting core database principles such as data independence, logical optimization, and cost-based plan enumeration.

Despite their expressive power, existing semantic database engines face a severe \textbf{scalability wall}. Traditional relational operators (e.g., hash joins, B-tree lookups, SIMD scans) execute at nanosecond-to-microsecond granularity~\cite{duckdb_2019, cstore_2005}. In stark contrast, invoking an autoregressive foundation LLM requires hundreds of milliseconds to several seconds per tuple, while incurring substantial monetary costs when querying cloud-hosted APIs~\cite{chen2023seed,  liu2024declarative, liu2025palimpchat, zhao2026larch,liskowski2026compositional}. Evaluating even a moderately sized analytical query over thousands or millions of tuples using full generative models quickly becomes computationally and financially expensive. While prior systems have explored various mitigation strategies---such as concurrent serving batching~\cite{lotus_2024, vllm}, physical model cascading~\cite{palimpzestCIDR, abacus}, prompt rewriting~\cite{docetl_2025}, or training lightweight proxy classifiers over precomputed embeddings~\cite{bigquery_proxy_2026, thalamusdb}---these techniques primarily optimize the scheduling of generative calls or trade execution time for heavy embedding storage and online training overheads, without resolving the underlying computational bottleneck.

We argue that the primary reason existing optimizations struggle to overcome this scalability wall is an underlying \textbf{architectural impedance mismatch}: \emph{semantic query engines are applying autoregressive, generative text decoders to evaluate non-generative, discrete relational predicates}. Traditional LLMs are architected to synthesize free-form natural language through sequential, token-by-token decoding. Yet, the overwhelming majority of relational query operators require strictly typed, deterministic decisions rather than conversational text: a semantic filter evaluates to a \textbf{Boolean} truth value; a semantic join condition evaluates \textbf{binary equivalence or compatibility}; a categorical projection selects from a discrete \textbf{Enum} of candidate classes; and a semantic ranking query requires a scalar or ordinal \textbf{Score}. Coercing a multi-billion-parameter autoregressive model to generate conversational output (e.g., generating words like ``\texttt{True}'' or verbose JSON structures) merely to parse and cast the result back into relational types is fundamentally inefficient. It consumes redundant GPU memory bandwidth, inflates latency by multiple orders of magnitude, and exposes the execution pipeline to syntax hallucination and parsing failures.

Recently, the machine learning community has witnessed the rise of non-generative, ``System One'' decision models---exemplified by TypeSafe AI's \textbf{Jev}~\cite{typesafe_jev_2026} alongside modern calibrated cross-encoders and discriminative foundation models. Rather than synthesizing text token by token through expensive autoregressive loops, these models evaluate an input state against a natural language specification in a single forward pass. They natively output calibrated probabilities across discrete decision primitives: binary Boolean judgments (e.g., \lstinline{Noul}), categorical selection over finite options (e.g., \lstinline{Choice}), and ordinal or continuous scoring (e.g., \lstinline{Score}). Trained with objectives tailored for calibrated decisions (such as RLCD) rather than conversational alignment (RLHF), these models deliver mathematically sound uncertainty metrics while achieving orders-of-magnitude lower latency and execution cost compared to frontier generative LLMs~\cite{jev_in_the_wild_2026}. While emerging applications have attempted to deploy such decision models within autonomous agentic workflows (e.g., agent routing and ticket dispatching)~\cite{jev_in_the_wild_2026}, agentic tasks frequently demand open-ended conversational generation and complex format synthesis---regimes where decision-only models struggle. Conversely, \textbf{relational semantic databases present an ideal match for decision models}. Relational queries require precisely the low-latency, typed determinations that these models natively supply, without requiring expensive embedding precomputation or online proxy model training.

In this paper, we present JEVDB%
\footnote{JEVDB stands for \textbf{J}ust \textbf{E}valuation and \textbf{V}erification \textbf{D}ata\textbf{b}ase. It is an independent academic research project developed by Purdue University, and is not affiliated with or endorsed by TypeSafe AI.},
a scalable semantic database system that bridges classical relational query theory with calibrated fast-path decision models. We name our system JEVDB to reflect its core design philosophy: \emph{Just Evaluation and Verification}. Rather than relying on expensive autoregressive text decoders for discrete relational predicates, JEVDB extracts \emph{just} the necessary attributes and performs \emph{just} the calibrated evaluation and verification required by relational algebra. Crucially, JEVDB is not coupled to any single proprietary backend; its architecture defines an adaptable, model-agnostic decision layer that unifies diverse non-generative architectures---including Jev, open-weight decision transformers, and fine-tuned cross-encoders---routing queries through high-throughput decision paths and selectively escalating to frontier generative LLMs only when deep contextual reasoning is required. Specifically, we make the following contributions:
\begin{itemize}[leftmargin=*]
    \item \textbf{Relational-Semantic Algebra Mapping:} We formalize the mapping of extended relational algebra operators ($\sigma_{\text{sem}}, \bowtie_{\text{sem}}, \Pi_{\text{sem}}, \Gamma_{\text{sem}}$) directly to native, single-pass decision primitives (\lstinline{Noul}, \lstinline{Choice}, \lstinline{Score}), eliminating token-generation overhead and syntax parsing from the semantic execution path.
    
    \item \textbf{Relational and Semantic Candidate Reduction:}
    To reduce the cost of large semantic joins, JEVDB combines two
    complementary forms of candidate reduction. Within explicit relational
    regions, it applies Yannakakis-style semijoin reduction to remove tuples
    that cannot participate in the relational query. Across latent semantic
    edges, JEVDB introduces \emph{Semantic Bloom Filters} (SBFs), which use
    registered necessary conditions to conservatively screen tuples or pairs
    before model evaluation. Extracted condition attributes can optionally be
    materialized in a reusable condition index and amortized across repeated
    queries.
    
    \item \textbf{Adaptive Three-Tier Execution Engine:} We design a hierarchical query execution framework:
    \begin{itemize}
        \item \emph{Tier 1 (Filter Fast-Path):} Relational predicate pushdown, Yannakakis-style semijoin reduction, and SBF screening.
        \item \emph{Tier 2 (Jev Decision Engine):} High-throughput parallel evaluation of semantic selections and joins using single-pass calibrated decision models.
        \item \emph{Tier 3 (LLM Slow-Path):} Selective fallback to generative frontier LLMs only when calibrated uncertainty exceeds an application-specified error threshold or open-ended text synthesis is explicitly requested.
    \end{itemize}

  \item \textbf{Evaluation on Diverse Benchmarks:} We evaluate JEVDB on both standard workloads in SemBench~\cite{lao2025sembench} and Shelob, a large-scale semantic join benchmark derived from TPC-DS. On SemBench, JEVDB achieves average latency speedups of $72.0\times$--$89.4\times$ (reaching up to hundreds of times faster on individual queries) and reduces costs by up to $18.7\times$, while matching or exceeding baseline quality on 18 of 21 queries. On Shelob joins with up to 540K candidates where existing baselines time out, JEVDB finishes every query with $95.7\%$--$97.5\%$ mean F1 while pruning $87.4\%$ of candidate pairs before inference, demonstrating scalable execution without compromising answer quality.
\end{itemize}

The remainder of this paper is structured as follows. Section~\ref{sec:related} reviews related work in detail. Section~\ref{sec:operators} formalizes our typed operator mappings, Semantic Bloom Filter design, and execution engine. Section~\ref{sec:eval} provides our empirical evaluation on SemBench and Shelob, and Section~\ref{sec:conclusion} concludes with future work discussion.

%% file: 02related.tex
\section{Related Work}
\label{sec:related}

JEVDB builds upon and synthesizes research across declarative semantic databases, algorithmic join reductions, and non-generative decision models.

\subsection{Semantic Database Systems and Scalability Approaches}
\label{sec:related:semantic-db}

Integrating machine learning and foundation models directly into database engines has emerged as a major research frontier to empower analysts with declarative semantic reasoning~\cite{ai4db_2021, llm_survey_2025}. Early efforts focused on in-database analytic libraries and tensor execution runtimes, such as MADlib~\cite{mad_12}, in-database deep learning operators~\cite{konstantinos_2020, rieger_2022, klabe_2023}, and relational matrix encodings for neural networks~\cite{Schüle2024}. 

With the emergence of Large Language Models, recent semantic database systems have explored incorporating natural language reasoning into query processing~\cite{palimpzestCIDR, lotus_2024, docetl_2025, flock_2025, evadb_2023,zhao2025access}. To confront the severe latency and monetary cost of foundation models, existing systems explore four primary scalability paradigms:
\begin{enumerate}[leftmargin=*]
    \item \textbf{Parallel Execution and Serving Batching:} Systems such as LOTUS~\cite{lotus_2024} extend Pandas dataframes with declarative semantic operators (\lstinline{sem_filter}, \lstinline{sem_join}, \lstinline{sem_agg}) and rely on high-throughput model serving engines (e.g., vLLM~\cite{vllm}) by grouping prompts and issuing concurrent requests. Flock~\cite{flock_2025} integrates scalar and aggregate LLM inference inside DuckDB~\cite{duckdb_2019}, while EvaDB~\cite{evadb_2023} and Aero~\cite{aero_2025} support adaptive predicate reordering over AI User-Defined Functions (UDFs). While batching improves GPU hardware saturation, it does not alter the fundamental FLOPs or token-by-token decoding latency required per tuple.
    
    \item \textbf{Cost-Based Model Cascades:} Palimpzest~\cite{palimpzestCIDR} introduces a declarative physical optimization scheme for unstructured data extraction. Its cost-based optimizer, Abacus~\cite{abacus}, searches a Pareto frontier over execution cost, latency, and quality by cascading queries through smaller language models before escalating uncertain cases to expensive frontier LLMs. However, even lightweight generative models incur sequential decoding overheads and require extensive prompt engineering across diverse model architectures.
    
    \item \textbf{Agentic Query Rewriting and Plan Decomposition:} DocETL~\cite{docetl_2025} addresses the complexity of document processing by employing an agentic query optimizer that uses an LLM to rewrite monolithic extraction prompts into modular map-reduce steps and entity resolution blocks. While effective for complex document pipelines, the rewritten plans still evaluate each sub-task using generative LLM calls.

    \item \textbf{Query Approximation and Lightweight Proxy Models:} To bypass full foundation models, approximate query processing (AQP) techniques have been developed for semantic workloads. ThalamusDB~\cite{thalamusdb} employs deterministic query approximation and user-guided sample labeling to accelerate multimodal natural language predicates. In industrial data warehouses, Google BigQuery and AlloyDB~\cite{bigquery_proxy_2026} recently demonstrated that lightweight linear models (e.g., Logistic Regression) trained over precomputed dense embeddings can achieve over $100\times$ speedups for semantic filters (\lstinline{AI.IF}) and semantic ranking (\lstinline{AI.RANK}).
\end{enumerate}

While these paradigms mitigate aspects of query overhead, they introduce substantial systems trade-offs. Generative engines (LOTUS, Palimpzest, DocETL) remain fundamentally bound by autoregressive token decoding for discrete relational outputs. Meanwhile, proxy models (BigQuery) trade execution time for heavy secondary storage bloat, severe cold-start sampling and fitting delays on ad-hoc queries, and accuracy collapse under class-imbalanced predicates. In contrast, JEVDB eliminates the generative bottleneck by compiling semantic operators directly into non-generative, single-pass decision primitives, requiring neither offline vector precomputation nor online model fitting.

\subsection{Semantic Joins and Algorithmic Reductions}
\label{sec:related:joins}

Evaluating semantic join predicates across large tables is notoriously challenging due to the quadratic $O(\vert{}L\vert{} \cdot \vert{}R\vert{})$ pairwise comparison space. Existing engines address this bottleneck primarily through heuristic pre-filtering. For example, LOTUS~\cite{lotus_2024} proposes the \emph{Map-Search-Filter} pattern, which projects tuples into dense embedding spaces and uses approximate nearest neighbor (ANN) vector retrieval to prune candidates down to top-$k$ pairs before invoking cross-encoder LLM verification. Alternatively,  ~\cite{decomp_join:2025} introduce featurized-decomposition joins, extracting structured attribute representations from unstructured texts to soundly prune pairwise combinations with accuracy guarantees.

However, embedding-based candidate retrieval introduces heavy index maintenance overheads and cannot eliminate the relational ``dangling tuple'' problem across multi-table query plans. JEVDB takes a fundamentally different, theoretically grounded approach: it couples operator decomposition with the classical Yannakakis algorithm~\cite{yannakakis}, executing bidirectional Bloom filter semijoin passes~\cite{rpt, rpt+} to eliminate non-participating tuples in linear time. Furthermore, JevDB's Semantic Bloom Filters (SBFs) use registered necessary conditions to conservatively screen candidates across latent semantic edges.

\subsection{Non-Generative Decision Models and the Jev Ecosystem}
\label{sec:related:jev}

Recent advances in machine learning highlight a growing divergence between ``System Two'' generative language models (optimized for multi-step reasoning traces and conversational generation) and fast ``System One'' decision models~\cite{typesafe_jev_2026}. TypeSafe AI introduced Jev~\cite{typesafe_jev_2026}, a non-generative model trained via Reinforcement Learning for Calibrated Decisions (RLCD). Instead of auto-regressively emitting tokens, Jev maps an input state and a natural language instruction to calibrated probabilities across discrete interfaces (\lstinline{Noul}, \lstinline{Choice}, and \lstinline{Score}).

A recent empirical study by ~\cite{jev_in_the_wild_2026} analyzed 2,170 open-source repositories and documented rapid adoption of Jev for agentic tasks, such as tool routing, ticket triage, and runtime guardrails across orchestration frameworks like LangChain and Vercel AI Gateway. However, their study revealed an inherent limitation: agentic workflows frequently demand multi-step conversational synthesis, code generation, and complex format restructuring---tasks where decision-only models cannot operate in isolation. 

In contrast to conversational agents, relational databases represent an ideal execution environment for Jev. Relational operators require deterministic, typed decisions (Booleans for selections and joins, Enums for classification, and numeric scores for ranking) rather than open-ended prose. JEVDB is the first system to formalize this relational-semantic synergy, embedding calibrated decision models directly into the database engine's core execution loop.

%% file: 03method.tex
\begin{figure}[!ht]
\centering
\includegraphics[width=0.80\textwidth]{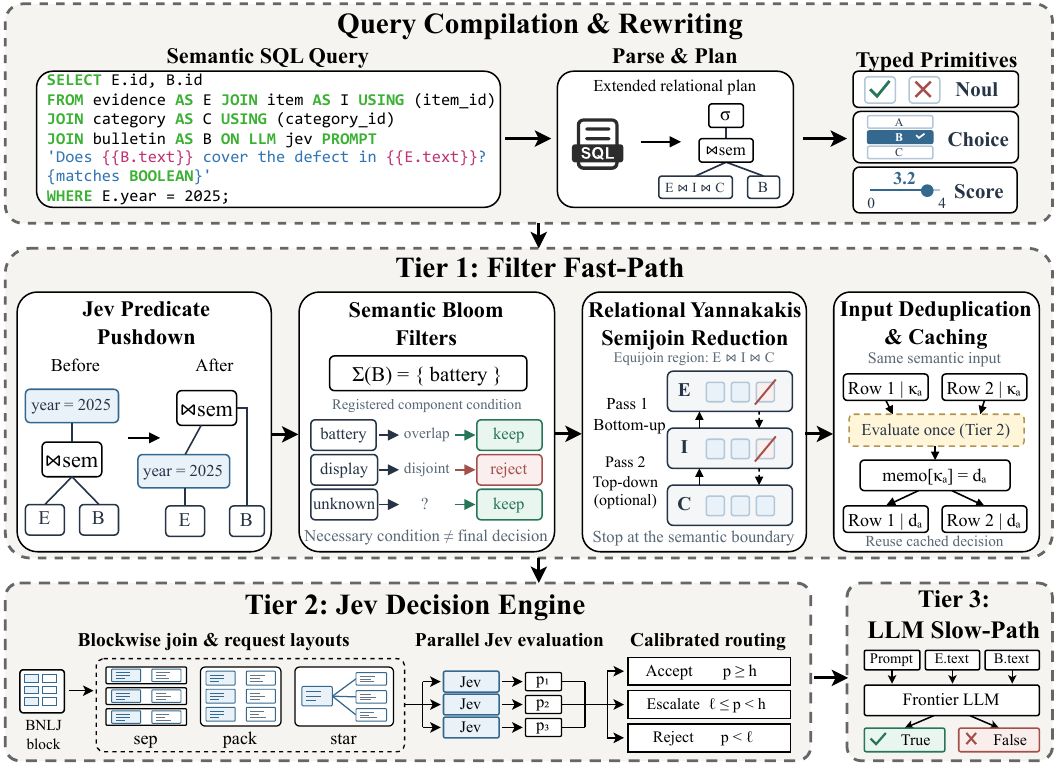}
\caption{JEVDB architecture and query execution workflow. Semantic SQL queries are compiled into relational plans and typed decision primitives. Tier~1 reduces candidates through relational predicate pushdown, Semantic Bloom Filters, relational Yannakakis semijoin reduction, and input deduplication. Tier~2 evaluates surviving inputs with the Jev decision engine, while uncertain decisions are escalated to a frontier LLM in Tier~3.}
\label{fig:architecture}
\end{figure}

\FloatBarrier

\section{Methodology}
\label{sec:method}
\label{sec:operators}

Figure~\ref{fig:architecture} illustrates the end-to-end architecture and query execution lifecycle of JEVDB. At a high level, JEVDB functions as an integrated relational-semantic engine that decouples fast-path discrete relational decisions from heavy generative model inference through three coordinated stages:
\begin{enumerate}[leftmargin=*]
    \item \textbf{Query Compilation and Rewriting:} Declarative SQL queries containing semantic expressions are parsed into extended relational plans. The engine isolates distinct text inputs to avoid redundant inferences and maps semantic tasks directly to typed decision contracts (\S\ref{sec:method:rewriting}).
    \item \textbf{Tier~1: Candidate Reduction:} 
    Prior to model evaluation, JEVDB pushes down standard relational filters, applies exact Yannakakis-style reduction within relational regions, and uses SBFs to screen candidates across semantic edges (\S\ref{sec:jevdb-candidate-reduction}).
    \item \textbf{Tier~2 \& Tier~3: Fast-Path Execution and Tiered Escalation:} Surviving candidates are batched using token-saving request layouts and evaluated by the Jev decision coprocessor in a single forward pass. Tuples with high calibrated confidence are immediately materialized into the columnar relational sink, while ambiguous edge cases within an uncertainty band $[\ell, h)$ are selectively escalated to a frontier generative LLM (\S\ref{sec:jevdb-join-execution}).
\end{enumerate}

To enable this pipeline, a decision model is registered once within the database catalog and subsequently referenced by name in relational queries. Listing~\ref{lst:jev-register} illustrates the registration syntax, specifying the target endpoint, batch size ($k$), concurrency ($n\_threads$), token budgets, and default confidence thresholds ($\tau$).

\begin{figure}[htbp]
\centering
\begin{lstlisting}[style=jevsql,float=false,firstnumber=1,nolol=true,
  xleftmargin={\dimexpr(\textwidth-166pt)/2\relax},
  xrightmargin={\dimexpr(\textwidth-166pt)/2\relax}]
CREATE LLM MODEL jev PATH 'jev-<version>'
ON PROMPT API '<endpoint>'
OPTIONS {'layout': '<layout>',
         'k': <batch_size>,
         'n_threads': <concurrency>,
         'budget_tokens': <token_budget>,
         'threshold': <threshold>};
\end{lstlisting}
\captionof{lstlisting}{Registering a Jev model and its execution options.}
\label{lst:jev-register}
\end{figure}

Queries invoke the registered model through an inline \lstinline{LLM} clause containing the natural language instruction, referenced column attributes, and a typed output declaration. For example, \lstinline|{{review_text}}| binds the input text from the current tuple, while \lstinline|{matches BOOLEAN}| declares a typed Boolean projection. JEVDB maps each relational semantic operator directly to the corresponding Jev primitive, returning strictly typed SQL values without intermediate JSON serialization or regex parsing (Table~\ref{tab:typed-decisions}).

In the following discussion, let $R$ denote an input relation, $x(t)$ the text projection of tuple $t \in R$, and $\phi$ the natural language task specification.

\begin{table}[htbp]
\centering
\small
\caption{Jev decision primitives and corresponding SQL result types.}
\label{tab:typed-decisions}
\begin{tabular}{lll}
\toprule
Semantic Operation & Jev Primitive & SQL Result Type \\
\midrule
Filter ($\sigma_{\text{sem}}$) & \lstinline{Noul} & \lstinline{BOOLEAN} / \lstinline{DOUBLE} \\
Classify ($\Pi_{\text{sem}}$) & \lstinline{Choice} & \lstinline{VARCHAR} \\
Ranking ($\Gamma_{\text{sem}}$ / $\tau_k$) & \lstinline{Score} & \lstinline{DOUBLE} \\
Join ($\bowtie_{\text{sem}}$) & \lstinline{Noul} & \lstinline{BOOLEAN} \\
\bottomrule
\end{tabular}
\end{table}

\subsection{Semantic Operator Rewriting via Jev Primitives}
\label{sec:method:rewriting}

JEVDB decomposes semantic SQL expressions into four foundational relational operators, replacing token-by-token autoregressive generation with single-pass decision inference.

\subsubsection{Semantic Filter ($\sigma_{\text{sem}}$)}
\label{sec:method:filter}
A semantic filter selects tuples satisfying a natural language predicate $\phi$. JEVDB evaluates the predicate via the \lstinline{Noul} binary decision primitive, obtaining the model's calibrated probability that $x(t)$ satisfies $\phi$~\cite{typesafe_noul_2026}:
\begin{equation}
    p_{\phi}(t) = \operatorname{Noul}(\phi, x(t)) \in [0,1].
\end{equation}
Given an application-defined decision threshold $\tau \in [0,1]$, the semantic selection operator is defined as:
\begin{equation}
    \sigma^{\phi,\tau}(R) = \{t \in R \mid p_{\phi}(t) \geq \tau\}.
\end{equation}
JEVDB returns a \lstinline{BOOLEAN} evaluation ($p_{\phi}(t) \ge \tau$) when declared in a predicate filter, or preserves the raw probability $p_{\phi}(t)$ as a \lstinline{DOUBLE} attribute (Listing~\ref{lst:jev-filter}). For semantic top-$k$ selections, JEVDB evaluates candidate probabilities, orders qualifying tuples by decreasing confidence, and emits the top-$k$ rows via relational \lstinline{LIMIT}.

\begin{figure}[htbp]
\centering
\begin{minipage}[t]{0.48\textwidth}
\centering
\small\textit{(a) Boolean predicate}\par
\begin{lstlisting}[style=jevsql,float=false,firstnumber=1,nolol=true,
  xleftmargin={\dimexpr(\textwidth-138pt)/2\relax},
  xrightmargin={\dimexpr(\textwidth-138pt)/2\relax}]
SELECT review_id,
  LLM jev PROMPT
    'The review is positive. '
    '{{review_text}} '
    '{matches BOOLEAN}' AS matches
FROM reviews;
\end{lstlisting}
\end{minipage}
\hfill
\begin{minipage}[t]{0.48\textwidth}
\centering
\small\textit{(b) Calibrated probability}\par
\begin{lstlisting}[style=jevsql,float=false,firstnumber=1,nolol=true,
  xleftmargin={\dimexpr(\textwidth-122pt)/2\relax},
  xrightmargin={\dimexpr(\textwidth-122pt)/2\relax}]
SELECT review_id,
  LLM jev PROMPT
    'The review is positive. '
    '{{review_text}} '
    '{p DOUBLE}' AS p
FROM reviews;
\end{lstlisting}
\end{minipage}
\captionof{lstlisting}{Semantic filter returning Boolean and calibrated probability values.}
\label{lst:jev-filter}
\end{figure}

\paragraph{Pushdown vs. Pull-up Trade-offs.}
In classical relational database systems, \emph{predicate pushdown} is a canonical heuristic optimization rule~\cite{hellerstein2019looking}: filters are pushed down as close to the base storage scan as possible to minimize the volume of intermediate tuples. However, when selection predicates involve model inference, this heuristic often breaks down. As observed in modern semantic data systems ~\cite{ipdb_2026, abacus}, foundation LLM calls are orders of magnitude more expensive than standard relational operators. In such engines, blindly pushing down expensive semantic predicates can severely degrade performance if tuples could have otherwise been cheaply eliminated by downstream relational filters or selective joins. Consequently, prior semantic systems frequently resort to \emph{predicate pull-up}, complex cost-benefit predicate reordering, or speculative routing~\cite{aero_2025, evadb_2023}.

In contrast, because Jev evaluates predicates in a single non-generative forward pass, Jev-based filters are consistently fast and cheap. This predictable, low-cost profile fundamentally simplifies query optimization: JEVDB can safely \emph{push down} semantic filters $\sigma^{\phi,\tau}$ directly to the scan and join input paths without incurring the severe execution penalties of generative LLMs. Restoring pushdown as a default execution strategy removes the need for fragile pull-up heuristics, reduces intermediate tuple materialization early in the query pipeline, and conforms directly to the standard rule engines of mature relational database systems~\cite{hellerstein2019looking, duckdb_2019}.

\subsubsection{Semantic Classification ($\Pi_{\text{sem}}$)}
\label{sec:method:classify}
Semantic classification projects each tuple onto a discrete label set $C=\{c_1,\ldots,c_m\}$ supplied in the query (e.g., classifying customer sentiment or support ticket categories). JEVDB maps this operation to Jev's \lstinline{Choice} primitive. Each candidate label can optionally include natural language descriptions clarifying decision boundaries. \lstinline{Choice} returns the label with the highest model probability~\cite{typesafe_choice_2026}:
\begin{equation}
    c_{\phi,C}(t) \in \underset{c \in C}{\operatorname{arg\,max}} \Pr(c \mid x(t),\phi,C).
\end{equation}
JEVDB exposes the selected label as a typed \lstinline{VARCHAR} column, functioning as a semantic projection that downstream operators can consume as grouping keys or join attributes. As shown in Listing~\ref{lst:jev-classify}, target labels can be declared directly in the SQL expression or registered globally in the model options.

\begin{figure}[htbp]
\centering
\begin{minipage}[t]{0.48\textwidth}
\centering
\small\textit{(a) Labels in SQL expression}\par
\begin{lstlisting}[style=jevsql,float=false,firstnumber=1,nolol=true,
  xleftmargin={\dimexpr(\textwidth-158pt)/2\relax},
  xrightmargin={\dimexpr(\textwidth-158pt)/2\relax}]
SELECT review_id,
  LLM jev PROMPT
    'Classify the sentiment. '
    '{{review_text}} '
    '{label VARCHAR["positive", '
                    '"negative", '
                    '"mixed"]}' AS label
FROM reviews;
\end{lstlisting}
\end{minipage}
\hfill
\begin{minipage}[t]{0.48\textwidth}
\centering
\small\textit{(b) Labels in model OPTIONS}\par
\begin{lstlisting}[style=jevsql,float=false,firstnumber=1,nolol=true,
  xleftmargin={\dimexpr(\textwidth-166pt)/2\relax},
  xrightmargin={\dimexpr(\textwidth-166pt)/2\relax}]
CREATE LLM MODEL jev PATH 'jev-<v>'
ON PROMPT API '<endpoint>'
OPTIONS {
  'choice_criteria':
    '{"positive":"positive",'
     '"negative":"negative",'
     '"mixed":"mixed"}'
};

SELECT review_id,
  LLM jev PROMPT
    'Classify the sentiment. '
    '{{review_text}} '
    '{label VARCHAR}' AS label
FROM reviews;
\end{lstlisting}
\end{minipage}
\captionof{lstlisting}{Semantic classification with candidate labels declared inline or in model options.}
\label{lst:jev-classify}
\end{figure}

\subsubsection{Semantic Ranking ($\Gamma_{\text{sem}}$ / $\tau_k$)}
\label{sec:method:ranking}
Semantic ranking evaluates tuples against an ordered sequence of qualitative levels $L=(\ell_0,\ldots,\ell_{m-1})$ (e.g., from \lstinline{Very negative} to \lstinline{Very positive}) and computes a continuous score. JEVDB evaluates these levels via Jev's \lstinline{Score} primitive. Letting $q_i(t)$ denote the model's calibrated probability mass assigned to level $\ell_i$, Jev computes the expected ordinal index~\cite{typesafe_score_2026}:
\begin{equation}
    s_{\phi,L}(t) = \sum_{i=0}^{m-1} i \cdot q_i(t), \qquad \text{where } \sum_{i=0}^{m-1} q_i(t) = 1.
\end{equation}
The resulting score $s_{\phi,L}(t) \in [0, m-1]$ is returned as a native \lstinline{DOUBLE} column (Listing~\ref{lst:jev-ranking}). Retaining the fractional component preserves fine-grained probability differences, enabling consistent sorting within SQL \lstinline{ORDER BY} clauses as well as downstream numeric aggregations (e.g., \lstinline{AVG(rating)}).

\begin{figure}[htbp]
\centering
\begin{minipage}[t]{0.48\textwidth}
\centering
\small\textit{(a) Inline ranking criteria}\par
\begin{lstlisting}[style=jevsql,float=false,firstnumber=1,nolol=true,
  xleftmargin={\dimexpr(\textwidth-162pt)/2\relax},
  xrightmargin={\dimexpr(\textwidth-162pt)/2\relax}]
SELECT review_id, movie_id,
  LLM jev PROMPT
    'Rate the review sentiment. '
    '{{review_text}} '
    '{s DOUBLE["Very negative", '
               '"Negative", "Neutral", '
               '"Positive", '
               '"Very positive"]}'
    AS rating
FROM reviews
ORDER BY rating DESC, review_id;
\end{lstlisting}
\end{minipage}
\hfill
\begin{minipage}[t]{0.48\textwidth}
\centering
\small\textit{(b) Registered ranking criteria}\par
\begin{lstlisting}[style=jevsql,float=false,firstnumber=1,nolol=true,
  xleftmargin={\dimexpr(\textwidth-166pt)/2\relax},
  xrightmargin={\dimexpr(\textwidth-166pt)/2\relax}]
CREATE LLM MODEL jev PATH 'jev-<v>'
ON PROMPT API '<endpoint>'
OPTIONS {
  'score_criteria':
    '["Very negative", "Negative", '
     '"Neutral", "Positive", '
     '"Very positive"]'
};

SELECT review_id, movie_id,
  LLM jev PROMPT
    'Rate the review sentiment. '
    '{{review_text}} '
    '{s DOUBLE}' AS rating
FROM reviews
ORDER BY rating DESC, review_id;
\end{lstlisting}
\end{minipage}
\captionof{lstlisting}{Semantic ranking returning continuous scores for relational sorting.}
\label{lst:jev-ranking}
\end{figure}

\subsubsection{Semantic Join ($\bowtie_{\text{sem}}$)}
\label{sec:jevdb-semantic-join}
A semantic join correlates records across relations $L$ and $R$ according to a cross-attribute natural language predicate $\phi$. JEVDB evaluates pairwise candidate tuples $(l, r) \in L \times R$ using \lstinline{Noul}:
\begin{equation}
  p_\phi(l,r) = \operatorname{Noul}(\phi, x(l), x(r)), \qquad l \in L,\; r \in R,
\end{equation}
retaining pairs where $p_\phi(l,r) \geq \tau$. Listing~\ref{lst:jev-join} demonstrates matching known-issue bulletins with return reports. Because an unconstrained semantic join demands $|L| \cdot |R|$ expensive model inferences, JEVDB applies candidate reduction prior to inference and executes remaining pairs using join-aware batching.

\begin{figure}[htbp]
\centering
\begin{lstlisting}[style=jevsql,float=false,firstnumber=1,nolol=true,
  xleftmargin={\dimexpr(\textwidth-166pt)/2\relax},
  xrightmargin={\dimexpr(\textwidth-166pt)/2\relax}]
SELECT i.issue_id, e.evidence_id
FROM known_issue i JOIN return_evidence e
  ON LLM jev PROMPT
    'Does this bulletin cover the defect '
    'described in this return report? '
    '{matches BOOLEAN} '
    '{{e.report_text}} {{i.bulletin_text}}';
\end{lstlisting}
\captionof{lstlisting}{Semantic join matching issue bulletins with customer reports.}
\label{lst:jev-join}
\end{figure}

\subsection{Candidate Reduction for Semantic Joins}
\label{sec:jevdb-candidate-reduction}

A semantic join exposes a \emph{latent} relation: the set of matching pairs is unknown until the semantic predicate is evaluated. JEVDB therefore reduces model work on both sides of this semantic boundary. Within ordinary relational regions, exact semijoin reduction removes tuples that cannot participate in the relational part of the query. Across a semantic edge, \emph{Semantic Bloom Filters} (SBFs) use registered necessary conditions to reject tuples or pairs that cannot satisfy the semantic predicate. Finally, JEVDB evaluates each distinct surviving semantic input once and reuses the resulting decision.

\subsubsection{Semantic Bloom Filters}
\label{sec:method:sbf}
Let $s(l,r) \in \{0,1\}$ denote the Boolean semantic predicate between $l \in L$ and $r \in R$. For a source relation $L$, define the semantic neighborhood that can still match on the target side as
\begin{equation}
    \mathcal{A}_s(L) = \{r \in R \mid \exists l \in L : s(l,r)=1\}.
\end{equation}
We call a summary $\Sigma_s(L)$ with probe function $\operatorname{Probe}(\Sigma_s(L),r) \in \{0,1\}$ a \emph{Semantic Bloom Filter} if a negative probe can be used to reject $r$. An SBF is \emph{lossless} when
\begin{equation}
    \operatorname{Probe}(\Sigma_s(L),r)=0
    \;\Longrightarrow\;
    r \notin \mathcal{A}_s(L).
\end{equation}
Thus, like a conventional Bloom filter, an SBF may retain false positives but must not remove a true match under its lossless contract. The difference is where the summary comes from. A relational Bloom filter summarizes keys of an \emph{explicit} join predicate such as $L.k=R.k$. An SBF summarizes a \emph{latent semantic relation}, whose matching pairs are unavailable without model inference, through a conservative condition implied by the semantic predicate. Its physical representation therefore need not be a Bloom bit array. JEVDB can use a bit vector, hash set, or other compact summary as long as the same one-sided probe contract is respected.

\textbf{Necessary conditions.}
A cheap predicate $P(l,r)$ can support an SBF when it is necessary for the semantic predicate:
\begin{equation}
    s(l,r)=1 \implies P(l,r)=1.
\end{equation}
The target set admitted by $P$,
\begin{equation}
    \widehat{\mathcal{A}}_P(L)
    = \{r \in R \mid \exists l \in L : P(l,r)=1\},
\end{equation}
contains the true semantic neighborhood $\mathcal{A}_s(L)$. Consequently, a target tuple outside $\widehat{\mathcal{A}}_P(L)$ can be discarded before the expensive semantic judgment. For example, if matching defect reports and engineering bulletins requires compatible components, component compatibility can screen obviously impossible candidates without deciding whether the two texts actually describe the same defect.

\textbf{From conditions to executable SBFs.}
Necessary conditions may be supplied by developers or proposed offline by decomposing the semantic predicate into conservative attribute-level requirements. JEVDB separates this discovery step from query execution: the optimizer only consumes conditions that have been registered for a particular semantic predicate and model configuration. Once a condition is registered, its required attributes can be extracted once per distinct document—for example, with Jev's \lstinline{Choice} primitive—and reused across candidate comparisons. 
JEVDB can materialize these reusable document-level attributes as a \emph{condition index}, which is shared across queries that use the same
registered condition and model configuration.

For equality-style conditions, let $K_L(l)$ and $K_R(r)$ denote possibly multi-valued extracted keys. JEVDB supports two physical realizations. A \emph{unary SBF} summarizes the source keys,
\begin{equation}
    \Sigma(L) = \bigcup_{l \in L} K_L(l),
\end{equation}
and rejects $r$ when $K_R(r) \cap \Sigma(L)=\emptyset$. This form can reduce an entire target relation before candidate pairs are materialized. A \emph{pairwise screen} instead rejects an individual pair $(l,r)$ when $K_L(l) \cap K_R(r)=\emptyset$, which is more selective but operates after pairs have been formed. Unknown keys are treated conservatively and never cause rejection.

\textbf{Correctness and cost.}
The SBF guarantee depends on the provenance of both $P$ and its extractors. If the necessary condition is sound and the required keys are extracted soundly, SBF pruning is lossless: a negative probe proves that no semantic match has been removed. If either the condition or the extractor is model-derived and only empirically validated, the SBF is an approximate filter and may trade recall for cost; JEVDB records this distinction in the registered SBF specification. This guarantee is separate from calibration of the downstream Jev probability.

JEVDB supports both unary SBFs and pairwise screens as physical realizations of a registered necessary condition. Detailed cost-based selection among these alternatives is left to future work.
Registered conditions and their associated extracted attributes may also be materialized and reused across queries. For repeated workloads over the same data and semantic condition, the one-time preparation cost is amortized, leaving only inexpensive screening before the surviving candidates are evaluated by the semantic predicate.

\subsubsection{Relational Yannakakis Reduction}
\label{sec:method:yannakakis}
SBFs reduce candidates across a semantic edge. Before crossing that edge, JEVDB also exploits exact relational structure. Semantic queries frequently contain local selections and ordinary equijoins alongside natural language predicates. For a supported acyclic inner-equijoin region, JEVDB roots the relational join tree at a relation incident to the semantic predicate and applies a Yannakakis-style semijoin schedule~\cite{yannakakis,Bernstein_reduction, rpt, rpt+}.

A bottom-up pass removes tuples that lack witnesses in descendant relations, reducing the boundary relation toward the tuples that can still participate in the relational join. An optional top-down pass propagates the reduced bindings back through the component to shrink the remaining relations as well. JEVDB realizes these semijoins with \emph{conventional relational Bloom filters} over explicit join keys. Because Bloom filters have no false negatives and the original exact join conditions remain in the plan, false positives only weaken pruning and do not change the query result.

This stage deliberately stops at the semantic boundary: it never infers which tuples match across $s(l,r)$. Relational Yannakakis reduction removes candidates using known relational constraints; SBFs are the complementary mechanism that transfers conservative information across the otherwise latent semantic edge.

\subsubsection{Input Deduplication and In-Query Caching}
\label{sec:method:memo}
After relational and semantic screening, the remaining candidate relation $C \subseteq L \times R$ can still contain duplicate semantic inputs~\cite{ipdb_2026}. Let $\kappa(l,r)$ denote the complete tuple of values referenced by the semantic expression, including text and scalar attributes. JEVDB evaluates the predicate once for each distinct input:
\begin{equation}
    U = \delta\!\left(\pi_{\kappa}(C)\right),
\end{equation}
and stores the resulting probability or Boolean decision in an in-query hash cache. The cached decision is then reattached to all relational tuples sharing the same semantic input, preserving SQL duplicate semantics while avoiding repeated model calls.

This optimization applies to pointwise semantic predicates whose decision is defined per input pair. Within a query, in-query caching ensures that repeated occurrences of the same input reuse the identical decision; operators whose semantics depend jointly on the composition of an entire candidate batch are not deduplicated in this way.


\subsection{Join-Aware Execution and Tiered Escalation}
\label{sec:jevdb-join-execution}

Once candidate reduction terminates, surviving pairs must be evaluated by the model backend. JEVDB optimizes this phase through payload-aware request layouts, vectorized block batching, and selective multi-tier escalation.

\subsubsection{Anchor-Sharing Request Layouts}
\label{sec:method:layouts}
Unlike traditional LLM endpoints that process solitary prompt strings, a single Jev API request can ingest a shared context state alongside multiple independent typed questions. JEVDB exploits this capability to reduce transport serialization and token footprint through three physical request layouts:
\begin{itemize}[leftmargin=*]
    \item \textbf{Separate (\texttt{sep}):} Dispatches each pair $(l_i, r_j)$ as an independent API request. This layout minimizes state coupling and is favored when candidate pairs are sparse or evenly distributed across partitions.
    \item \textbf{Packed (\texttt{pack}):} Bundles multiple independent string pairs $\{(l_1, r_1), (l_2, r_2), \ldots\}$ into a single network payload. This amortizes HTTP connection latency and packet header overhead when candidate groups are small.
    \item \textbf{Star (\texttt{star}):} Designates a high-degree tuple as a shared anchor record (e.g., a single engineering bulletin $l_i$) and attaches candidate counterparts $\{r_1, \ldots, r_m\}$ as distinct questions within the same request.
\end{itemize}
The \texttt{star} layout provides substantial efficiency gains: the anchor text $x(l_i)$ is included only once in the model input, reducing token transfer by up to $(m-1) \cdot |x(l_i)|$ tokens. JEVDB's execution engine dynamically selects between \texttt{star} and \texttt{pack} based on the average candidate degree per anchor tuple, bounded by the model's maximum prompt context budget.

\subsubsection{Block-Level Vectorized Batching}
\label{sec:method:batching}
Relational database engines typically execute in vectorized batches (e.g., 2,048 tuples per chunk) to maximize CPU cache locality. However, evaluating semantic predicates tuple-at-a-time or vector-at-a-time exposes insufficient concurrency to fully saturate cloud model serving backends.

JEVDB resolves this throughput impedance mismatch inside a specialized \emph{Blockwise Nested-Loop Join} (BNLJ) operator. The join operator buffers candidate pairs across an entire memory-resident block of outer tuples before performing model evaluation. It partitions uncached candidate pairs into optimal request layouts and dispatches them across an asynchronous worker pool of $n\_threads$ concurrent requests. Once the batch inference returns, a second local scan matches join keys against the populated judgment buffer, emitting qualifying rows in native columnar chunks. This design decouples the high concurrency demanded by the model backend from the engine's internal vector pipelining.

Our block-level execution shares the motivation of recent semantic block nested-loop joins~\cite{trummer2025implementing}, but differs in the model interface and unit of batching. Prior work places blocks from both inputs into a generative prompt and asks the LLM to enumerate matching pairs. JEVDB instead retains candidate enumeration in the database and batches explicit typed decisions, using shared-state layouts and an in-query memo over the surviving candidate pairs.

\subsubsection{Calibrated Three-Tier LLM Escalation}
\label{sec:method:escalation}
While Jev provides rapid, low-cost decisions for standard predicates, complex queries may occasionally encounter ambiguous corner cases requiring deep contextual reasoning. JEVDB addresses this trade-off through an adaptive, three-tier execution hierarchy governed by dual confidence thresholds $[\ell, h]$ (where $0 \le \ell \le h \le 1$):
\begin{equation}
    \operatorname{Eval}(l, r) = 
    \begin{cases}
        \text{True}, & \text{if } p_\phi(l, r) \geq h \quad \text{(Tier 2: Jev Fast-Accept)}, \\
        \text{False}, & \text{if } p_\phi(l, r) < \ell \quad \text{(Tier 2: Jev Fast-Reject)}, \\
        \operatorname{LLM}(\phi, x(l), x(r)), & \text{if } p_\phi(l, r) \in [\ell, h) \quad \text{(Tier 3: Generative Escalation)}.
    \end{cases}
\end{equation}

Tuples exhibiting high model confidence are accepted or rejected immediately by Jev (Tier 2). Only ambiguous pairs falling within the uncertainty band $[\ell, h)$ are escalated to an expensive frontier generative LLM (Tier 3). 

To configure thresholds $\ell$ and $h$ without requiring manual user tuning, JEVDB employs an automated online calibration protocol:
\begin{enumerate}[leftmargin=*]
    \item During query initialization, JEVDB routes a small uniform sample of candidate pairs (e.g., 200 pairs) directly to the frontier LLM judge.
    \item It evaluates Jev's calibrated probabilities $p_\phi(l, r)$ against the LLM's binary determinations on this sample.
    \item It computes the tightest uncertainty band $[\ell, h)$ such that the expected recall loss $\Delta_{\text{rec}}$ and precision loss $\Delta_{\text{prec}}$ remain strictly within user-specified quality tolerances:
    \begin{equation}
        \Delta_{\text{rec}} \le \epsilon_{\text{recall}}, \qquad \Delta_{\text{prec}} \le \epsilon_{\text{precision}}.
    \end{equation}
\end{enumerate}
This routing keeps confident candidates on the decision-model fast path while reserving the reasoning model for candidates within the calibrated uncertainty band.

\subsubsection{Condition-index scoring}
Registered semantic conditions can also be evaluated approximately from the reusable condition index. For each document, the index stores the decision model's probability distribution over the attributes required by a registered condition. JEVDB combines the corresponding document-level distributions to obtain a lightweight pair score, which can replace a direct pairwise Jev call as the first stage of the cascade. This shifts the first-stage inference cost from the number of candidate pairs to the number of distinct documents; the index is constructed once and reused across queries.

%% file: 04eval_tmp.tex
\section{Evaluation}
\label{sec:eval}

We evaluate JEVDB on two complementary benchmarks. First, SemBench provides a common basis for comparison with existing semantic query engines on public workloads and commonly used scale factors. Second, Shelob evaluates semantic joins over relations derived from TPC-DS, with substantially larger candidate spaces than those exercised by current semantic benchmarks. We focus on end-to-end performance: rather than isolating individual optimizations, we report the monetary cost, query latency, and answer quality of complete query executions.

\subsection{Experimental Setup}
\label{sec:eval-setup}

\textbf{Systems.}
We compare JEVDB with LOTUS~\cite{lotus_2024}, Palimpzest~\cite{palimpzestCIDR}, and ThalamusDB~\cite{thalamusdb}. The SemBench baselines use
Gemini~2.5 Flash and the configurations used by SemBench~\cite{lao2025sembench}. For SemBench, we use the numbers published by the
benchmark. For Shelob, we run LOTUS~1.1.3, Palimpzest~0.8.2, and ThalamusDB~0.1.15 with \texttt{gpt-6-luna} and SemBench-style
configurations.

\textbf{JEVDB configurations.}
We expose two end-to-end execution modes. \textbf{JEVDB-Flash} is the latency-oriented mode: it evaluates semantic predicates
with Jev (\texttt{jev-1.13.0}) at a fixed 0.5 threshold and does not invoke a reasoning model. Filters and joins use
\texttt{noul}, classification uses \texttt{choice}, and ranking uses \texttt{score}. \textbf{JEVDB} is the quality-controlled
mode used in the Shelob comparison. It retains Jev as the fast path, applies pairwise SBF screening, and selectively escalates
uncertain candidates to \texttt{gpt-6-luna}. The cascade calibrates 400 items per predicate; $\epsilon_r$ and $\epsilon_p$
denote the tolerated recall and precision loss relative to the calibration judge and are both set to 0.05. Confident Jev
decisions remain on the fast path, so the reasoning model is used only when needed to meet the configured quality target.
The second stage packs 16 items per request and uses 64 concurrent requests.

JEVDB additionally materialize a condition index for registered predicates and reuse its document-level semantic
information across queries. This preprocessing is optional: users can execute directly from the online Jev fast path, or build
the index once when repeated queries make its cost worth amortizing. For the reported Shelob experiments, the index covers
20,558 documents and costs 15\,s and \$0.54 of Jev inference to construct. We exclude this one-time cost from individual query
costs and report it separately. The XS measurements predate index reuse and repeat document extraction (\$0.063 total, already included), S and M reuse the stored index for all five Shelob queries.

\textbf{SemBench.}
We evaluate all 21 queries from SemBench's Table~4 that operate on text and structured attributes, spanning filters (F), joins (J), classification (C), and ranking (R) alongside standard relational operators and \texttt{LIMIT} (L). These queries cover four benchmark domains: Movie, E-Commerce, MMQA, and Cars, evaluated under the official per-scenario scale factors (Table~3). We exclude SemBench queries that require raw multi-modal inputs (e.g., image perception), leaving multi-modal decision execution as future work. Baseline metrics are the published SemBench numbers using Gemini~2.5 Flash (averaged over five runs); four queries are omitted for ThalamusDB due to lack of engine support. JEVDB-Flash executes each query once using 20 concurrent HTTP workers, matching the baseline concurrency setting.

We follow SemBench's official quality metric for each query: F1 for retrieval and join queries (including the benchmark's
rules for \texttt{LIMIT}); $1/(1+\text{relative error})$ for Movie Q3, Q4, Q8 and Cars Q4; adjusted Rand index for
E-Commerce Q5; macro-F1 for Cars Q10; and Spearman correlation for Movie Q9 and Q10.

\textbf{Shelob Dataset.}
We build Shelob to complement SemBench with semantic joins over relations at analytical-benchmark scale. SemBench~\cite{lao2025sembench} offers broad coverage of existing semantic operators, whereas Shelob derives its relational backbone from a conventional analytical benchmark. It starts from TPC-DS SF1 and augments the database with textual return reports, known-issue bulletins,
and release-note documents linked to the underlying relational entities. The full SF1 universe contains 503,588 return reports,
and a report--bulletin join exposes 18.1 million candidate pairs before JEVDB-specific semantic reduction.

For head-to-head comparison, we evaluate five Shelob queries on three increasing tiers, XS, S, and M. Q1--Q4 exercise direct
semantic joins between reports, bulletins, and release notes under different predicates; Q5 additionally restricts
report--release candidates through an ordinary item-lineage join. Across Q1--Q5, the aggregate candidate space grows from
81,528 pairs at XS to 332,262 at S and 1.37 million at M, while the largest individual M-tier join contains 540K candidates.
Table~\ref{tab:shelob-workload} summarizes the workload scale; the full-SF1 row is included only to show the scale of the
underlying workload universe.

\begin{table*}[htbp]
\centering
\caption{Shelob workload scale. Each tier contains 36 known-issue bulletins. Candidate counts are measured after
query-defined relational restrictions but before JEVDB-specific semantic reduction.}
\label{tab:shelob-workload}
\small
\setlength{\tabcolsep}{8pt}
\begin{tabular}{@{}lrrrrr@{}}
\toprule
Tier
& Return reports
& Release notes
& Q1/Q4
& Q2/Q3
& Q5 \\
& ($|R|$)
& ($|N|$)
& $|R| \times |B|$
& $|B| \times |N|$
& lineage-constrained \\
\midrule
XS
& 609
& 511
& 21,924
& 18,396
& 888 \\
S
& 3,029
& 1,523
& 109,044
& 54,828
& 4,518 \\
M
& 15,005
& 3,672
& 540,180
& 132,192
& 22,652 \\
\midrule
Full SF1$^\ast$
& 503,588
& 6,668
& 18.1M
& 240,048
& 756,416 \\
\bottomrule
\end{tabular}

\vspace{2pt}
\end{table*}

\textbf{Shelob protocol and metrics.}
All systems execute the same logical Shelob queries end-to-end over
the corresponding tier data using their native query interfaces; no
semantic candidate relation is materialized in advance. All runs use a
1,200\,s execution limit. For completed queries, we report F1 against
Shelob's reference answers. For runs that reach the timeout, we do not
report F1 because the answer is incomplete; instead, Table~\ref{tab:shelob-main}
reports progress as the fraction of candidate pairs evaluated before
termination. The reported model cost of a timed-out run is the amount
incurred before the stop. JEVDB-Flash uses the \texttt{pack} layout with $k=100$ and 64 concurrent workers on Shelob; JEVDB uses the quality-controlled configuration above. Quality is F1
against Shelob's reference answers. We report F1 on the full output; ambiguous
(\emph{unknown}) references count as neither true nor false positives.

\textbf{Cost and environment.}
We compute JEVDB cost from measured token counts and list prices at the time of the runs: Jev at \$0.042 per million input
tokens and \texttt{gpt-6-luna} at \$0.10/\$0.50 per million input/output tokens. For SemBench, we use the published baseline
costs. For Shelob, we compute baseline costs from API-reported token usage at the \texttt{gpt-6-luna} rates above,
with cached input at \$0.01 and cache writes at \$0.125 per million tokens. Latency is end-to-end wall-clock query time, including relational processing and model requests; the reported full JEVDB
runs additionally include loading the tier task tables and writing the result. LOTUS embeddings are computed on CPU in our experiments; GPU acceleration could reduce embedding overhead and improve query latency.
 

\subsection{SemBench: Performance on Existing Semantic Workloads}
\label{sec:eval-sembench}

\begin{table*}[htbp]
\centering
\caption{SemBench: cost (USD), quality and latency (s) per query. Baseline numbers are from SemBench's Table~4 (Gemini 2.5
Flash, mean of five runs); JEVDB-Flash is a single run. Bold: best per metric and query (quality compared at two decimals).
Op: F filter, J join, C classification, R ranking, L \texttt{LIMIT}. $\dagger$: run through a Python driver.
--: not supported. }
\label{tab:sembench}
\scriptsize
\setlength{\tabcolsep}{3.2pt}
\begin{tabular}{@{}ll rrr rrr rrr rrr@{}}
\toprule
 & & \multicolumn{3}{c}{JEVDB-Flash} & \multicolumn{3}{c}{LOTUS} & \multicolumn{3}{c}{Palimpzest} & \multicolumn{3}{c}{ThalamusDB} \\
\cmidrule(lr){3-5} \cmidrule(lr){6-8} \cmidrule(lr){9-11} \cmidrule(l){12-14}
Query & Op & Cost & Qual. & Lat. & Cost & Qual. & Lat. & Cost & Qual. & Lat. & Cost & Qual. & Lat. \\
\midrule
\multicolumn{14}{@{}l}{\textit{Movie}}\\
Q1 & F\,L & 0.0052 & \textbf{1.00} & \textbf{0.51} & 0.09 & \textbf{1.00} & 33.1 & 0.001 & \textbf{1.00} & 3.8 & \textbf{0.0004} & 0.95 & 4.2 \\
Q2 & F\,L & \textbf{0.00032} & \textbf{1.00} & \textbf{0.38} & 0.01 & \textbf{1.00} & 2.1 & 0.01 & \textbf{1.00} & 29.7 & 0.002 & 0.92 & 1.9 \\
Q3 & F & \textbf{0.00032} & \textbf{0.88} & \textbf{0.37} & 0.005 & 0.64 & 2.1 & 0.01 & 0.64 & 4.6 & 0.002 & 0.74 & 3.1 \\
Q4 & F & \textbf{0.00032} & \textbf{0.88} & \textbf{0.42} & 0.005 & 0.64 & 2.8 & 0.02 & 0.74 & 4.4 & 0.002 & 0.74 & 3.8 \\
Q5$^\dagger$ & J\,L & \textbf{0.00011} & 0.90 & \textbf{1.11} & 2.38 & 0.59 & 536.5 & 0.01 & 0.39 & 1.9 & 0.001 & \textbf{1.00} & 2.3 \\
Q6$^\dagger$ & J\,L & \textbf{0.00011} & 0.80 & \textbf{0.28} & 1.81 & 0.67 & 432.4 & 0.01 & 0.83 & 2.3 & 0.0009 & \textbf{0.84} & 1.7 \\
Q7$^\dagger$ & J & \textbf{0.00050} & \textbf{0.82} & \textbf{1.17} & 1.81 & 0.21 & 431.8 & 7.72 & 0.68 & 1056.1 & 0.15 & 0.57 & 636.9 \\
Q8 & C & \textbf{0.00061} & \textbf{0.96} & \textbf{0.41} & 0.004 & 0.93 & 2.3 & 0.02 & 0.86 & 4.3 & 0.005 & 0.83 & 6.8 \\
Q9 & R & \textbf{0.0011} & \textbf{0.78} & \textbf{0.63} & 0.02 & 0.75 & 4.9 & 0.05 & \textbf{0.78} & 5.7 & -- & -- & -- \\
Q10 & R & \textbf{0.015} & 0.41 & \textbf{0.93} & 0.13 & 0.40 & 30.9 & 0.38 & \textbf{0.42} & 39.2 & -- & -- & -- \\
\midrule
\multicolumn{14}{@{}l}{\textit{E-Commerce}}\\
Q1 & F & \textbf{0.0050} & \textbf{1.00} & \textbf{0.65} & 0.06 & \textbf{1.00} & 12.2 & 0.08 & \textbf{1.00} & 12.2 & 0.03 & \textbf{1.00} & 34.8 \\
Q5 & C & \textbf{0.0051} & 0.98 & \textbf{0.66} & 0.04 & \textbf{0.99} & 7.4 & 0.07 & 0.98 & 6.6 & -- & -- & -- \\
Q7 & J & 0.0913 & \textbf{0.92} & \textbf{2.35} & 1.33 & 0.75 & 199.4 & 1.79 & \textbf{0.92} & 287.6 & \textbf{0.08} & 0.51 & 97.7 \\
\midrule
\multicolumn{14}{@{}l}{\textit{MMQA}}\\
Q3a & F & \textbf{0.0013} & 0.81 & \textbf{0.47} & 0.01 & \textbf{0.83} & 4.2 & 0.02 & 0.80 & 4.5 & 0.01 & 0.75 & 11.0 \\
Q3f & F & \textbf{0.0013} & 0.86 & \textbf{0.47} & 0.01 & \textbf{1.00} & 4.2 & 0.02 & \textbf{1.00} & 8.0 & 0.01 & \textbf{1.00} & 7.0 \\
Q6a & F & \textbf{0.00042} & \textbf{1.00} & \textbf{0.36} & 0.01 & \textbf{1.00} & 4.4 & 0.02 & \textbf{1.00} & 4.0 & 0.002 & 0.33 & 5.6 \\
Q6b & F & \textbf{0.00042} & \textbf{1.00} & \textbf{0.36} & 0.01 & \textbf{1.00} & 3.8 & 0.02 & \textbf{1.00} & 4.1 & 0.002 & \textbf{1.00} & 5.5 \\
Q6c & F & \textbf{0.00042} & \textbf{1.00} & \textbf{0.41} & 0.01 & \textbf{1.00} & 4.6 & 0.02 & \textbf{1.00} & 4.7 & 0.002 & 0.53 & 13.4 \\
\midrule
\multicolumn{14}{@{}l}{\textit{Cars}}\\
Q1 & F & \textbf{0.18} & \textbf{0.91} & \textbf{7.77} & 1.74 & 0.90 & 550.0 & 2.44 & 0.69 & 465.6 & 1.37 & 0.81 & 829.7 \\
Q4 & F & \textbf{0.18} & 0.99 & \textbf{7.72} & 1.71 & 0.99 & 822.0 & 2.41 & 0.99 & 443.6 & 1.34 & \textbf{1.00} & 768.8 \\
Q10 & C & \textbf{0.57} & \textbf{0.55} & \textbf{14.1} & 3.09 & 0.41 & 618.1 & 4.69 & 0.51 & 594.9 & -- & -- & -- \\
\bottomrule
\end{tabular}
\end{table*}

Table~\ref{tab:sembench} shows that the decision-model fast path is already competitive on the workloads for which current
semantic systems are designed. JEVDB-Flash is the fastest system on all 21 queries and the cheapest on 19. Across the complete
21-query set, JEVDB-Flash takes 41.50\,s and costs \$1.06, compared with 3,709\,s and \$14.28 for LOTUS and 2,988\,s and
\$19.81 for Palimpzest. Thus, over this workload as a whole, JEVDB-Flash reduces aggregate latency by about 89.4$\times$ and
72.0$\times$, and aggregate model cost by about 13.5$\times$ and 18.7$\times$, relative to LOTUS and Palimpzest, respectively.
ThalamusDB supports 17 of the 21 queries; its published runs total 2,434\,s and \$3.01, and JEVDB-Flash is faster on every
supported query.
\begin{figure*}[htbp]
    \centering
    \includegraphics[width=0.82\textwidth]{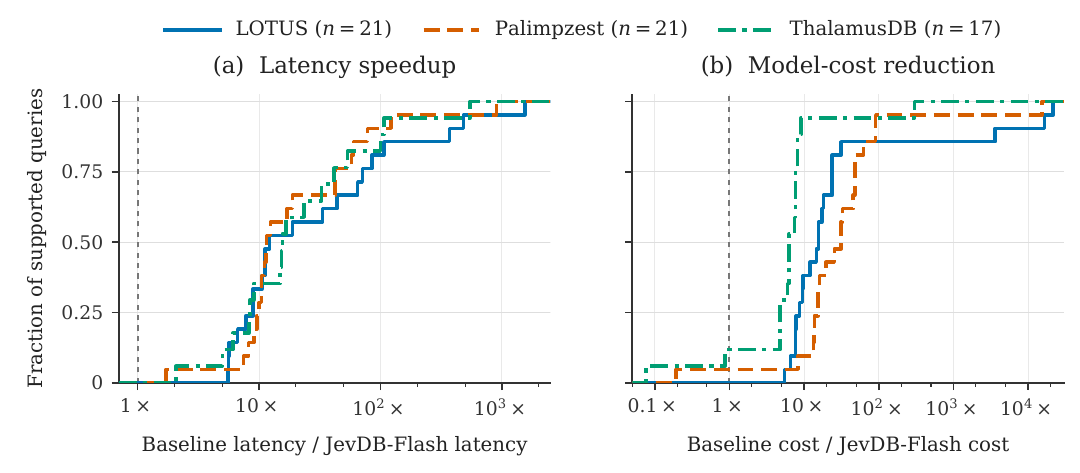}
    \caption{Distribution of per-query improvements of JEVDB-Flash over
    SemBench baselines. The $x$-axis reports the ratio of baseline latency
    or model cost to JEVDB-Flash; values above $1\times$ favor JEVDB-Flash.
    Each curve is normalized over the queries supported by the corresponding
    baseline (21 for LOTUS and Palimpzest, 17 for ThalamusDB).}
    \label{fig:sembench-ecdf}
\end{figure*}
These efficiency gains do not come from accepting uniformly worse answers. At the two-decimal precision of the published
results, JEVDB-Flash is best or tied for best quality on 14 of the 21 queries and is within 0.02 of the best system on four more. The
largest deficits are Movie Q5 (0.90 versus 1.00), Movie Q6 (0.80 versus 0.84), and MMQA Q3f (0.86 versus 1.00). The first two are
\texttt{LIMIT 10} joins, where one or two returned pairs materially change the score. Conversely, JEVDB gains substantially on
Movie Q7 (0.82 versus 0.21--0.68) and Movie Q3/Q4 (0.88 versus 0.64--0.74).

Figure~\ref{fig:sembench-ecdf} shows that the efficiency gains are not
driven by a small number of outlier queries. JEVDB-Flash is faster than
all three baselines across their supported queries, with most latency
speedups exceeding an order of magnitude. Model-cost reductions are also
broadly distributed across the workload, although a small number of queries
remain cheaper with the baseline systems.



SemBench shows that typed decision models can replace chat-style generation for a broad range of existing semantic operators while preserving comparable answer quality and substantially reducing latency and monetary cost. Shelob complements this result by explicitly scaling semantic-join candidate spaces over a TPC-DS-derived analytical workload: Q1/Q4 grow from 21,924 candidate pairs at XS to 540,180 at M, while the full SF1 workload contains 18.1 million such pairs.

\subsection{Shelob: Scaling Semantic Joins over Analytical Data}
\label{sec:eval-shelob}

\begin{table*}[htbp]
\centering
\caption{Shelob XS/S/M end-to-end results for JEVDB and existing systems:
F1 (\%) for completed runs, model cost (USD), and query latency (s).
Bold marks the best F1 among completed runs in each row.
$^T$ and gray shading denote a timeout under the 1200\,s limit; its cost reflects only work completed before termination.
For timed-out runs, Qual.\ shows progress (percentage of candidate pairs evaluated).}
\label{tab:shelob-main}
\scriptsize
\setlength{\tabcolsep}{2.8pt}
\usetikzlibrary{patterns}
\definecolor{shelobtimeoutbg}{RGB}{216,219,222}
\newcommand{\shelobtimeout}[1]{%
  \tikz[baseline=(value.base)]{%
    \node[preaction={fill=shelobtimeoutbg,draw=none},
      pattern=dots,pattern color=white,rounded corners=1.5pt,
      inner xsep=1.5pt,inner ysep=0.9pt,outer sep=0pt] (value) {\strut #1};}%
}
\begin{tabular}{@{}ll rrr rrr rrr rrr rrr@{}}
\toprule
 & & \multicolumn{3}{c}{JEVDB}
 & \multicolumn{3}{c}{JEVDB-Flash}
 & \multicolumn{3}{c}{LOTUS}
 & \multicolumn{3}{c}{Palimpzest}
 & \multicolumn{3}{c}{ThalamusDB} \\
\cmidrule(lr){3-5}
\cmidrule(lr){6-8}
\cmidrule(lr){9-11}
\cmidrule(lr){12-14}
\cmidrule(l){15-17}
Tier & Query
 & Qual. & Cost & Lat.
 & Qual. & Cost & Lat.
 & Qual. & Cost & Lat.
 & Qual. & Cost & Lat.
 & Qual. & Cost & Lat. \\
\midrule

XS & Q1
 & \textbf{96.7} & 0.089 & 73.7
 & 88.9 & 0.150 & 2.07
 & 77.1 & 1.492 & 1{,}171.30
 & 93.7 & 4.601 & 817.13
 & \shelobtimeout{(31.0\%)} & \shelobtimeout{0.076} & \shelobtimeout{1{,}200$^T$} \\

 & Q2
 & \textbf{98.1} & 0.069 & 66.5
 & 71.7 & 0.129 & 1.71
 & 79.6 & 1.924 & 1{,}193.12
 & 94.0 & 3.962 & 705.11
 & \shelobtimeout{(30.5\%)} & \shelobtimeout{0.075} & \shelobtimeout{1{,}200$^T$} \\

 & Q3
 & \textbf{97.2} & 0.136 & 91.8
 & 73.8 & 0.129 & 3.85
 & 79.0 & 2.063 & 1{,}168.93
 & 93.6 & 4.047 & 726.65
 & \shelobtimeout{(26.6\%)} & \shelobtimeout{0.075} & \shelobtimeout{1{,}200$^T$} \\

 & Q4
 & 98.1 & 0.058 & 45.5
 & 30.4 & 0.150 & 4.24
 & 64.4 & 1.951 & 1{,}196.85
 & \textbf{98.4} & 4.450 & 801.14
 & \shelobtimeout{(27.8\%)} & \shelobtimeout{0.078} & \shelobtimeout{1{,}200$^T$} \\

 & Q5
 & \textbf{97.4} & 0.021 & 43.1
 & 66.7 & 0.006 & 0.500
 & \textbf{97.4} & 0.218 & 69.54
 & \textbf{97.4} & 0.216 & 77.58
 & \textbf{97.4} & 0.192 & 233.37 \\

\midrule

S & Q1
 & \textbf{94.2} & 0.431 & 266
 & 86.8 & 0.749 & 22.5
 & \shelobtimeout{(0.0\%)} & \shelobtimeout{0.000} & \shelobtimeout{1{,}200$^T$}
 & \shelobtimeout{(30.3\%)} & \shelobtimeout{7.115} & \shelobtimeout{1{,}200$^T$}
 & \shelobtimeout{(7.1\%)} & \shelobtimeout{0.083} & \shelobtimeout{1{,}200$^T$} \\

 & Q2
 & \textbf{96.8} & 0.094 & 77.7
 & 73.2 & 0.385 & 9.79
 & \shelobtimeout{(0.0\%)} & \shelobtimeout{1.236} & \shelobtimeout{1{,}200$^T$}
 & \shelobtimeout{(58.8\%)} & \shelobtimeout{6.979} & \shelobtimeout{1{,}200$^T$}
 & \shelobtimeout{(9.1\%)} & \shelobtimeout{0.075} & \shelobtimeout{1{,}200$^T$} \\

 & Q3
 & \textbf{95.1} & 0.240 & 147
 & 71.7 & 0.385 & 9.53
 & \shelobtimeout{(0.0\%)} & \shelobtimeout{0.986} & \shelobtimeout{1{,}200$^T$}
 & \shelobtimeout{(57.1\%)} & \shelobtimeout{6.978} & \shelobtimeout{1{,}200$^T$}
 & \shelobtimeout{(8.4\%)} & \shelobtimeout{0.074} & \shelobtimeout{1{,}200$^T$} \\

 & Q4
 & \textbf{98.2} & 0.335 & 146
 & 27.0 & 0.748 & 21.1
 & \shelobtimeout{(0.0\%)} & \shelobtimeout{0.000} & \shelobtimeout{1{,}200$^T$}
 & \shelobtimeout{(31.2\%)} & \shelobtimeout{7.158} & \shelobtimeout{1{,}200$^T$}
 & \shelobtimeout{(5.8\%)} & \shelobtimeout{0.077} & \shelobtimeout{1{,}200$^T$} \\

 & Q5
 & \textbf{96.4} & 0.062 & 91.1
 & 61.3 & 0.031 & 1.81
 & 91.3 & 1.138 & 283.89
 & 93.3 & 1.124 & 233.21
 & 93.8 & 1.015 & 937.77 \\

\midrule

M & Q1
 & \textbf{96.3} & 2.534 & 1{,}125
 & 85.3 & 3.709 & 114
 & \shelobtimeout{(0.0\%)} & \shelobtimeout{0.000} & \shelobtimeout{1{,}200$^T$}
 & \shelobtimeout{(5.8\%)} & \shelobtimeout{6.715} & \shelobtimeout{1{,}200$^T$}
 & \shelobtimeout{(1.1\%)} & \shelobtimeout{0.069} & \shelobtimeout{1{,}200$^T$} \\

 & Q2
 & \textbf{94.9} & 0.262 & 172
 & 73.4 & 0.928 & 26.1
 & \shelobtimeout{(0.0\%)} & \shelobtimeout{0.000} & \shelobtimeout{1{,}200$^T$}
 & \shelobtimeout{(23.6\%)} & \shelobtimeout{6.353} & \shelobtimeout{1{,}200$^T$}
 & \shelobtimeout{(3.6\%)} & \shelobtimeout{0.070} & \shelobtimeout{1{,}200$^T$} \\

 & Q3
 & \textbf{91.5} & 0.592 & 275
 & 73.8 & 0.928 & 26.2
 & \shelobtimeout{(0.0\%)} & \shelobtimeout{0.000} & \shelobtimeout{1{,}200$^T$}
 & \shelobtimeout{(15.2\%)} & \shelobtimeout{4.698} & \shelobtimeout{1{,}200$^T$}
 & \shelobtimeout{(3.0\%)} & \shelobtimeout{0.065} & \shelobtimeout{1{,}200$^T$} \\

 & Q4
 & \textbf{98.7} & 2.175 & 883
 & 23.8 & 3.706 & 114
 & \shelobtimeout{(0.0\%)} & \shelobtimeout{0.000} & \shelobtimeout{1{,}200$^T$}
 & \shelobtimeout{(5.5\%)} & \shelobtimeout{6.349} & \shelobtimeout{1{,}200$^T$}
 & \shelobtimeout{(1.0\%)} & \shelobtimeout{0.073} & \shelobtimeout{1{,}200$^T$} \\

 & Q5
 & \textbf{96.9} & 0.245 & 161
 & 59.8 & 0.154 & 8.60
 & \shelobtimeout{(83.7\%)} & \shelobtimeout{4.872} & \shelobtimeout{1{,}200$^T$}
 & \shelobtimeout{(95.1\%)} & \shelobtimeout{5.349} & \shelobtimeout{1{,}200$^T$}
 & \shelobtimeout{(29.6\%)} & \shelobtimeout{1.434} & \shelobtimeout{1{,}200$^T$} \\

\bottomrule
\end{tabular}
\end{table*}

Table~\ref{tab:shelob-main} shows how end-to-end performance changes as the Shelob workload scales. Baseline completion
degrades rapidly with scale: only 14 of the 45 baseline tier--query runs finish within the 1,200\,s limit, and none finishes
at M. Timed-out entries report progress rather than F1, because they do not produce a complete answer. Q5 is the main
small-scale exception: its lineage restriction leaves only 888 candidates at XS, so all systems finish and the performance
gap is correspondingly less pronounced; the difference becomes clearer as Q5 grows to 4,518 and 22,652 candidates at S and M.

JEVDB-Flash isolates the latency-oriented decision-model path. It completes all 15 Shelob runs in 12.38\,s, 64.69\,s,
and 288.55\,s in total for XS, S, and M, at costs of \$0.56, \$2.30, and \$9.42. Its quality, however, depends strongly on
the predicate: Q1 reaches 85.3--88.9\% F1 and Q2/Q3 reach 71.7--73.8\%, while Q4 remains at only 23.8--30.4\%.
Mean F1 is 66.3\%, 64.0\%, and 63.2\% from XS to M. JEVDB-Flash therefore provides a useful low-latency option when Jev is
sufficient for the application, but a fixed-threshold decision does not provide uniformly high quality across predicates.

Full JEVDB takes the complementary choice: it preserves the Jev fast path but uses SBF screening, reusable semantic state,
and selective reasoning-model escalation to control answer quality. It completes all 15 runs with 91.5--98.7\% F1, with mean
F1 of 97.5\%, 96.1\%, and 95.7\% at XS, S, and M. This additional quality control increases latency relative to
JEVDB-Flash---320.6\,s, 727.8\,s, and 2,616\,s in total across the three tiers---but it does not require every candidate to
visit the reasoning model: confident Jev decisions remain on the fast path. In applications where the calibration stage
indicates that Jev already meets the target, execution can remain entirely on the fast path.

The optional condition index provides a second trade-off for repeated workloads. Its one-time construction cost can be
amortized across queries, allowing document-level semantic information to be reused rather than recomputed for every candidate pair. In the reported runs, the total model cost of full JEVDB is \$0.37, \$1.16, and \$5.81 across XS, S, and M, excluding the one-time \$0.54 index construction.

Across Q1--Q5 and all three Shelob tiers, SBF screening removes 87.4\% of the
1.78M candidate pairs before semantic evaluation, leaving 224K pairs.
Among these survivors, condition-index scoring reduces the number of candidates
escalated to the reasoning model from 194K to 87K, a further 55.2\% reduction.
Together, these optimizations substantially reduce both pairwise semantic
evaluation and expensive reasoning-model calls.

All JevDB and JevDB-Flash runs finish within the common 1,200\,s
limit. In contrast, no baseline completes any M-tier query. For timed-out
baseline runs, Table~\ref{tab:shelob-main} therefore reports progress
and the model cost incurred before termination rather than a completed query result.
LOTUS runs reporting 0\% progress reached the 1200\,s limit during embedding computation or threshold learning for filtering.

Overall, Shelob provides a controlled scaling test for semantic joins over TPC-DS-derived data. As the data and semantic candidate spaces grow, the baseline systems increasingly fail to complete the end-to-end queries within the execution
limit. In contrast, both JEVDB modes complete every XS/S/M query: JEVDB-Flash prioritizes latency, while full JEVDB sustains
95.7--97.5\% mean F1 with selective reasoning-model escalation. The largest individual measured semantic join reaches
540K candidates.

%% file: 05conclusion_zhengle.tex
\section{Conclusion and Future Work}
\label{sec:conclusion}

In this paper, we introduced JEVDB, a semantic database system that places
typed, non-generative decision primitives on the fast path for bounded
relational decisions. JEVDB maps semantic filters, joins, classification, and
ranking to \lstinline{Noul}, \lstinline{Choice}, and \lstinline{Score}, avoiding
autoregressive decoding and free-form output parsing whenever a direct decision
is sufficient. For semantic joins, JEVDB further reduces model work before
inference: exact Yannakakis-style semijoin reduction exploits explicit
relational structure, while \emph{Semantic Bloom Filters} (SBFs) use registered
necessary conditions to screen candidates across latent semantic edges.
Uncertain decisions can be selectively escalated to a reasoning model, and
reusable condition information can be materialized once and amortized across
repeated queries.

Our end-to-end evaluation covers two complementary benchmarks. On SemBench,
JEVDB-Flash achieves the lowest latency on all 21 evaluated queries and the
lowest cost on 19. Across the complete 21-query workload, it reduces aggregate
latency by about $88\times$ and $71\times$, and aggregate model cost by about
$12.5\times$ and $17.4\times$, relative to LOTUS and Palimpzest, respectively.
It achieves the best or tied-best quality on 14 queries and comes within 0.02
of the best result on four more. Shelob evaluates semantic joins over larger
TPC-DS-derived relations. Across Q1--Q5 and the XS, S, and M tiers, only 14 of
45 baseline runs complete within the 1,200\,s execution limit, and none
completes at M. Both JEVDB modes complete all 15 runs: JEVDB-Flash prioritizes
latency, while full JEVDB maintains arithmetic-mean full-output F1 of 97.5\%,
96.1\%, and 95.7\% from XS through M, with the largest evaluated semantic join
containing 540K candidate pairs. Together, these results show two complementary
ways to scale semantic query processing: typed decision models reduce the cost
of individual semantic decisions, while database-level pruning and selective
escalation reduce how much expensive reasoning is required.

\paragraph{Future Work.}
The current prototype leaves several opportunities for deeper integration
between semantic decisions and database systems.

\begin{itemize}[leftmargin=*]
    \item \textbf{Semantic-aware query optimization.}
    Cheap typed decisions make query-time sampling of semantic predicates much
    more practical. We plan to use sampled selectivities, candidate degrees,
    and observed escalation rates to estimate semantic cardinalities and guide
    join ordering, screening choices, and request layouts. These estimates can
    also be refined during execution, enabling adaptive replanning when observed
    semantic selectivities differ from the optimizer's initial assumptions
    \cite{avnur2000eddies}.

    \item \textbf{Reusable semantic state.}
    The condition index used by JEVDB suggests a broader opportunity to
    materialize semantic interpretations as persistent database state.
    Extracted categorical attributes, score distributions, and registered
    necessary conditions could support reusable secondary indexes and
    coarse-grained data skipping, amortizing semantic interpretation across
    repeated queries \cite{liu2023deep,liu2025data}. A corresponding systems
    challenge is maintenance: these summaries must be refreshed as data changes
    and versioned against the model, prompt, and registered condition on which
    they depend.

    \item \textbf{Scale-out semantic joins.}
    Necessary conditions can also provide routing information for distributed
    semantic joins. Rather than repartitioning only on exact equality keys,
    JEVDB could use semantic blocking keys to colocate likely matches while
    retaining selective replication or fallback evaluation for candidates that
    cannot be routed safely. We plan to study how such routing interacts with
    skew, approximation, and the cost of semantic inference.

    \item \textbf{Multi-modal semantic queries.}
    We plan to extend JEVDB's typed decision interface to predicates over text,
    images, and other modalities. A natural direction is to combine
    multi-modal representation encoders with non-generative decision heads so
    that cross-modal predicates, such as matching textual reports with
    inspection images, can be evaluated without autoregressive text generation.
    This also raises new questions for SBFs: which cross-modal necessary
    conditions can be extracted cheaply, reused safely, and pushed into
    candidate pruning?
\end{itemize}